\pdfoutput=1
\documentclass{iau}
\usepackage{graphicx}
\usepackage{multirow}
\title[IRDIS, testing the performances in laboratory] %% give here short title %%
{IRDIS, the dual-band imager camera of SPHERE: testing the performances in laboratory}

\author[Zurlo A. et al.]   %% give here short author list %%
{A. Zurlo$^1$, A. Vigan$^1$, C. Moutou$^1$, D. Mesa$^2$, R. Gratton$^2$, M. Langlois$^3$, J.-L. Beuzit$^4$, A. Costille$^1$, S. Desidera$^2$, K. Dolhen$^1$, C. Gry$^1$, F. Madec$^1$, D. Le Mignant$^1$, D. Mouillet$^4$ \and J.-F. Sauvage$^4$}

\affiliation{$^1$Aix Marseille Universit\'e, CNRS, LAM (Laboratoire d'Astrophysique de Marseille) UMR 7326, 13388, Marseille, France \\[\affilskip]
$^2$INAF - Osservatorio Astronomico di Padova - Vicolo dell'Osservatorio 5, I-35122, Padova, Italy \\[\affilskip]
$^3$CRAL, UMR 5574, CNRS, Universit\'e Lyon 1, 9 avenue Charles Andr\'e, 69561 Saint Genis Laval Cedex, France \\[\affilskip]

$^4$Institut de Planetologie et d'Astrophysique de Grenoble, UJF, CNRS, 414 rue de la piscine, 38400, Saint Martin d'Heres, France \\[\affilskip]}

\pubyear{2013}
\volume{299}  %% insert here IAU Symposium No.
\pagerange{119--201}
\jname{Exploring the formation and evolution of planetary systems}
\editors{A.C. Editor, B.D. Editor \& C.E. Editor, eds.}
\begin{document}

\maketitle

\begin{abstract}
Next year the second generation instrument SPHERE will begin science operations at the Very Large Telecope (ESO). This instrument will be dedicated to the search for exoplanets through the direct imaging techniques, with the new generation extreme adaptive optics. In this poster, we present the performances of one of the focal instruments, the Infra-Red Dual-beam Imaging and Spectroscopy (IRDIS). All the results have been obtained with tests in laboratory, simulating the observing conditions in Paranal.
We tested several configurations using the sub-system Integral Field Spectrograph (IFS) in parallel and simulating long coronographic exposures on a star, calibrating instrumental ghosts, checking the performance of the adaptive optics system and reducing data with the consortium pipeline. The contrast one can reach with IRDIS is of the order of 2$\times$10$^{-6}$ at 0.5 arcsec separation from the central star. 
\end{abstract}

\firstsection % if your document starts with a section,
\section{Test description}
IRDIS sub-system is one of the three focal sub-systems of SPHERE, working in different configurations in the range of 0.9--2.3 $\mu$m: dual-band imaging, long slit spectroscopy and polarimetric imaging.
IRDIS can observe in its dual-band imaging mode in parallel with the other focal sub-instrument IFS. A series of laboratory tests has been performed to find the final capabilities of the instrument and check possible issues.
Contrast plots of the NIRSUR (IRDIS in H2H3 and IFS in YJ) configuration is shown in Figure 1. The NIRSUR mode is the one optimized for planets search. It uses the spectral differential imaging (SDI) procedure (\cite[Racine et al. 1999]{Racine99}), a technique that permits the detection of faint companions in two narrow-band filters in which the contrast between primary and planet is very different (e.g. wavelength close to the methane band head). 

\begin{table}
  \begin{center}
  \caption{Observing conditions and data reduction.}
  \label{tab1}
 \scriptsize
  \begin{tabular}{l|l||l}\hline 
\multicolumn{3}{c}{{\bf Observing}  {\bf conditions}  \&         {\bf Data reduction}}\\ \hline 
Turbulence &  0.$^{''}$6 and 0.$^{''}$8 seeing & Backgroung subtraction + flat-filtering\\ 
Filter & Dual-band H2H3 & Rescaling\\
Coronograph & Apodized Lyot & Median recombination of the data-cube frames\\
Dithering & 4$\times$4 & Spectral subtraction (SDI)\\ 
Exposure time & 3584 s & NO ADI \\ 
IFS & yes & Injection of synthetic planets of different position angles and separation \\ \hline     
  \end{tabular}
 \end{center}
\vspace{1mm}
\end{table}

\begin{figure}[b]
% \vspace*{-2.0 cm}
\begin{center}
 \includegraphics[width=3.9in]{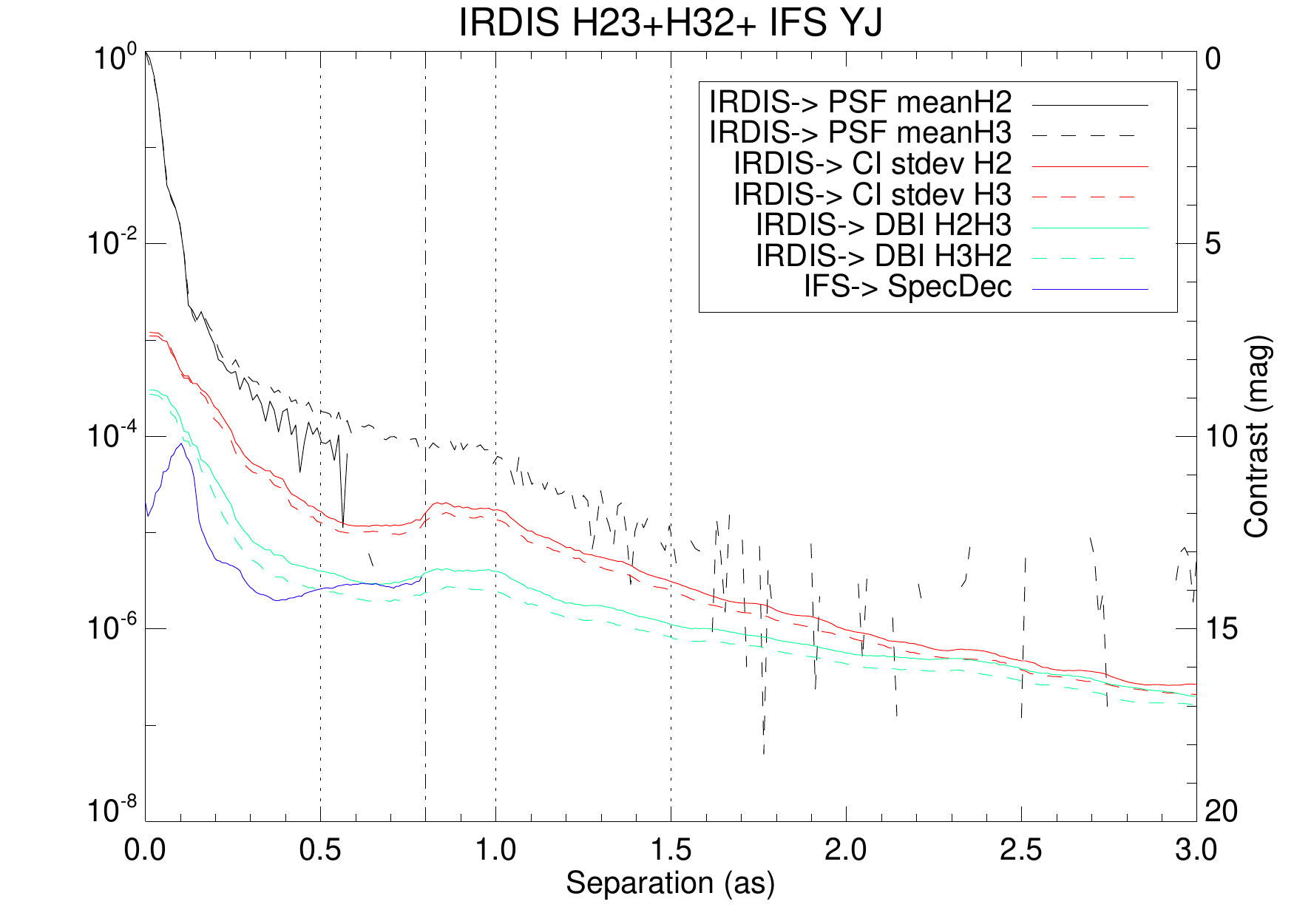} 
% \vspace*{-1.0 cm}
 \caption{Contrast plot showing the performances of IRDIS and IFS. The dual band filter H2-H3 has been used for IRDIS while IFS worked in YJ mode. The turbulence of 0.$^{''}$85 seeing. The plot shows the contrast with respect to the host star (black line) vs the separation in arcsec. Apodized Lyot coronographic profile (at 1 $\sigma$) is showed by the red lines, while in green is represented the result of the dual band imaging (DBI). The blue curve represents the IFS result after spectral deconvolution. Vertical lines indicates separation from the host star of 0.5, 0.8 (limit of IFS), 1.0 and 1.5 arcsec.}
   \label{fig1}
\end{center}
\end{figure}

\section{Conclusions}
The dual-band imager IRDIS will allow the detection of planets with a luminosity contrast of 2$\times$10$^{-6}$ at 0.5 arcsec of separation from the central star.  Synthetic fake planets have been injected in the raw data to better estimate the detection limits. We found that for planets with a contrast of 1$\times$10$^{-4}$ the signal to noise ratio (SNR) goes from 6.4 for internal to 33.3 for external planets. During the observations at the VLT the exposures will be acquired in pupil-stabilized mode, where the field will rotate, to take advantage of the ADI (\cite[Marois et al. 2005]{Marois05}) technique  in addition to the SDI. This configuration has not been tested realistically in the laboratory. Simulations predict a gain of a factor 10 in contrast when using ADI in addition to SDI. The photometric and astrometric characterization of faint companions, down to 17 mag of contrast (at 0.5 arcsec) with respect to their host stars, will be possible using the ADI and SDI techniques.

\end{document}